\documentclass[twocolumn,showpacs,superscriptaddress,amsmath,amssymb]{revtex4}
\usepackage[dvips]{graphicx}
\usepackage{dcolumn}
\usepackage{bm}
\usepackage{amsfonts}
\usepackage{amsmath}
\usepackage{amssymb}
\usepackage[utf8]{inputenc}
\usepackage{xcolor}
\usepackage{graphicx}
\usepackage{epsfig}
\usepackage{subfigure}
\usepackage[normalem]{ulem}
\usepackage{soul,xcolor}
\usepackage{ragged2e}
\usepackage{lineno}
\usepackage{hyperref}
\usepackage{wrapfig,lipsum,booktabs}

\begin{document}

%\preprint{APS/123-QED}

\title{Spin Squeezing via One-Axis Twisting in a Quadrupolar NMR system under relaxation effects}

\author{Adriane Consuelo Leal Auccaise} \email{adrianeleal25@gmail.com}
\affiliation{Universidade Estadual de Ponta Grossa, Departamento de F\'{\i}sica, CEP
84030-900, Ponta Grossa, PR, Brazil.}
\author{Antonio Sérgio Magalhães de Castro} 
\affiliation{Universidade Estadual de Ponta Grossa, Departamento de F\'{\i}sica, CEP
84030-900, Ponta Grossa, PR, Brazil.}

\begin{abstract}
This study investigates spin squeezed states in nuclear magnetic resonance (NMR) quadrupolar systems with spins $I=3/2$ and $I=7/2$ at room temperature, taking into account the effects of relaxation on the dynamics. The origin of spin squeezing is attributed to the interaction between the nuclear quadrupole moment and the electric field gradients in the molecular environment. The formal description of the nonlinear operators responsible for spin squeezing is achieved using the spin angular momentum representation via the one-axis twisting mechanism. This approach provides a framework for quantum control and metrology over the spin squeezing process while accounting for the influence of relaxation phenomena. Non-Cartesian angular momentum operators are proposed, with their variances products catching the quantum effects during the dynamics. An upper bound for the squeezing parameter and the Heisenberg uncertainty at thermal equilibrium are also predicted for any spin quantum number.
\end{abstract}


%\pacs{Valid PACS appear here}

\maketitle

\section{Introduction}\label{sec:level1}

Spin squeezing is a fundamental concept in quantum mechanics with broad applications in quantum technologies, including quantum computing, quantum control, quantum sensing, entanglement generation, and precision metrology \cite{kitagawa1991, wineland1992, kitagawa1993, Geremia2004, grond2009, Ma2011, PhysRevLett.123.260505.2019, Bao2020, PhysRevA.107.023724.2023}. It provides a key resource for quantum-enhanced measurements, enabling sensitivities that surpass the standard quantum limit and approach the Heisenberg limit. However, the practical implementation of spin squeezing is constrained by decoherence mechanisms, which arise from environmental interactions, spontaneous emission, and other noise sources. Understanding the interplay between spin squeezing and decoherence is thus crucial to advance quantum-enhanced metrology and quantum information processing.

The impact of decoherence on spin squeezing has been extensively investigated in various physical systems. Notable examples include cold atomic ensembles manipulated by Ramsey spectroscopy \cite{ulam2001spin}, spin-$1/2$ ensembles subject to different decoherence channels \cite{Wang2010}, spin networks coupled to a critical quantum environment \cite{sun2011spin}, and Bose–Einstein condensates \cite{AliceSinatra2012}. Additional studies have examined cold-confined atoms that interact dispersively with a paraxial laser beam, where both coherent scattering and decoherence due to diffuse scattering are considered \cite{baragiola2014three}. Further investigations have explored long-lived dipolar atomic arrays undergoing collective emission \cite{Tucker.102.051701.2020} and optomechanical systems coupled to single photons \cite{zhang2021single}. These studies provide insights into how decoherence affects quantum correlations and imposes limits on the performance of quantum-enhanced protocols.

Since real physical systems are inherently subject to relaxation effects, a detailed understanding of decoherence is necessary for developing robust quantum control techniques. In this work, we analyze the impact of relaxation-induced decoherence on spin squeezing, focusing on its implications for quantum control strategies. The evolution of the elements of the density matrix is described using the master equations that accounts for the dissipation of coherences and populations due to relaxation processes. Specifically, we study the dynamics of spin squeezing in a one-axis twisting model composed of quadrupolar spin-$3/2$ and spin-$7/2$ particles coupled to a Markovian environment in high-temperature regime. By characterizing the effects of relaxation on spin squeezing, our results provide a quantitative framework for assessing the feasibility of quantum-enhanced metrology under realistic conditions. This study contributes to the broader understanding of spin squeezing in dissipative quantum systems and offers insights relevant to experimental implementations in nuclear magnetic resonance.

This work is structured as follows. Sections \ref{Squeezing} and \ref{Relaxation} provide a brief introduction to the formalisms of spin squeezing and relaxation, respectively. The relationship between these concepts is explored in Section \ref{simulation}. In Section \ref{uncertainty}, the Heisenberg uncertainty is analyzed within the spin framework, and a new pair of angular momentum operators is proposed, with their uncertainty relation catching quantum coherences throughout the system dynamics. Section \ref{upper} establishes an upper bound for the squeezing parameter, macroscopicity measurement and the Heisenberg uncertainty at thermal equilibrium for arbitrary spin quantum numbers. Finally, our main conclusions are summarized in Section \ref{Conclusions}.

\section{Spin Squeezing and Schr\"odinger's cat state} \label{Squeezing}
In our study, the quadrupolar Hamiltonian is capable of generating a squeezed nuclear spin state by the one-axis twisting mechanism \cite{CoryCSS2003, auccaisePRL2015}, as well as the Schr\"odinger's cat state through free evolution dynamics \cite{consuelo2022nmr}. The Hamiltonian is described according to the NMR formalism of a quadrupolar system $I>1/2$ and the eigenstates are compatible with the magnetic quantum number $m=I, I-1, \cdots, -I+1, -I$. We write the NMR Hamiltonian in the laboratory frame, described in terms of two energy contributions given by
\begin{equation}
\boldsymbol{\mathcal{H}}_{NMR}=-\hbar(\omega_{L}-\omega_{rf})\boldsymbol{I}_{z}+\hbar\frac{\omega_{Q}}{6}(3\boldsymbol{I}_{z}^{2}-\boldsymbol{I}^{2})\text{.} \label{HNMR}
\end{equation}
The first energy contribution is due to the Zeeman energy, which describes the interaction between the nuclear magnetic moment, given by $-\hbar \gamma (\boldsymbol{I}_{x}, \boldsymbol{I}_{y}, \boldsymbol{I}_{z})$, and a strong static magnetic field $\boldsymbol{B}_{0}$ along the $z$ axis. Here, $\hbar$ is the reduced Planck's constant and ${B}_{0}$ represents the intensity of the static magnetic field, respectively.  The second energy contribution describes the interaction between the quadrupole moment of the nuclei and an electric field gradient around the nucleus,  considering an anisotropy oriented in the $z$-axis, and where $\omega_{Q}$ is the quadrupolar angular frequency.  

Assuming the resonance condition $\omega_{rf}=\omega_{L}$, where $\omega_{L}=\gamma B_{0}$ is the angular Larmor frequency and $\gamma$ is the gyromagnetic ratio of the nuclei, the NMR Hamiltonian in Eq.(\ref{HNMR}) corresponds to the one-axis twisting model of spin squeezing \cite{jin2007PRL} written as
\begin{equation}
\boldsymbol{\mathcal{H}}_{NMR}^{SS}=\hbar\frac{\omega_{Q}}{2}\boldsymbol{I}_{z}^{2}\text{,} \label{HNMRSS}
\end{equation}
where the term proportional to the quadratic spin operator, known as the Casimir energy contribution, $-\hbar\frac{\omega_{Q}}{6}\boldsymbol{I}^{2}$, is a constant factor and does not influence the generation of spin squeezing. 
 
In addition to generating spin squeezing, the Hamiltonian in Eq.(\ref{HNMRSS}) can also generate a Schr\"odinger cat state within the framework of the atom–field interaction scenario \cite{Aagarwal1981}. Specifically, the cat state emerges through the free evolution dynamics at time $\tau=\frac{\pi}{\omega_{Q}}$, during which the system undergoes a unitary transformation that produces a quantum superposition of spin coherent states, commonly referred to as a Schr\"odinger's cat state \cite{consuelo2022nmr}.

With the Hamiltonian established, the next step is to define the quantum state of the system. In this case, it is represented by a notation analogous to the well-known spin coherent state but adapted to be compatible with a nuclear spin system
\begin{equation}
\vert \zeta(\theta,\phi)\rangle=\sum_{m=I}^{-I} \frac{\zeta^{I+m}}{(1+\zeta^{*}\zeta)^{I}}\sqrt{ \frac{(2I)!}{(I+m)!(I-m)!}}\vert I,m \rangle \text{,} 
\end{equation}
where $\zeta=\tan \frac{\theta}{2}\exp(-i\phi)$ denotes the excitation parameter with $0 \leq \theta \leq \pi$ and $0 \leq \phi \leq 2\pi$. The spin coherent state $\vert \psi(0)\rangle=\vert \zeta(\pi/2,0)\rangle$ is selected as the initial state and corresponds to a symmetric Wigner quasi-probability distribution \cite{Aagarwal1981, benedict1999, sanchez-soto2013} in the spherical phase space around the $x$ axis.  
After transforming, the spin coherent state suffers also a unitary transformation in terms of the Hamiltonian of Eq.(\ref{HNMRSS}) and the quantum state achieves the squeezing shape of the distribution probability on the $yz$ plane, now described on a rotated $y'z'$ basis \cite{fernholz2008}.

To evaluate the degree of squeezing, we monitor the squeezing parameter expressed as \cite{law2001,jin2007PRA,jin2007PRL, auccaisePRL2015}
\begin{equation}
\xi=\frac{\left( \Delta\boldsymbol{I}_{\mathbf{n}}\right) _{\min}}{\sqrt{I/2}}\text{,}
\end{equation}
where $\left( \Delta\boldsymbol{I}_{\mathbf{n}}\right) _{\min}$ represents the minimum variance of a spin component $\boldsymbol{I}_{\mathbf{n}}$ orthogonal to the mean spin $\left\langle \boldsymbol{I}\right\rangle $. Explicitly,
\begin{eqnarray}
\xi & = & \frac{\sqrt{\frac{1}{2}C-\frac{1}{2}\sqrt{A^{2}+B^{2}}}}{\sqrt{I/2}} \text{,} \label{ParametroSqueezing}   
\end{eqnarray}
where $A=\left\langle \boldsymbol{I}_{z}^{2}-\boldsymbol{I}_{y}^{2}\right\rangle $, $B=\left\langle \boldsymbol{I}_{z}\boldsymbol{I}_{y}+\boldsymbol{I}_{y}\boldsymbol{I}_{z}\right\rangle $ and $C=\left\langle \boldsymbol{I}_{z}^{2}+\boldsymbol{I}_{y}^{2}\right\rangle $ are combinations of sums and products of the spin operators $\boldsymbol{I}_{z}$ and $\boldsymbol{I}_{y}$, depending on the orientation of a vector $\mathbf{n}=(1,0,0)$ which is aligned along the $x$-axis in a tridimensional space of the angular momentum operator mean values. In addition, we have the angle $\alpha_{\xi} = \frac{1}{2}\arctan \left( B / A \right)$ giving the geometrical orientation of the squeezing \cite{jin2007PRA,jin2007PRL}.

The spin squeezing parameter and the angle $\alpha_{\xi}$ are determined by combinations of the values $A$, $B$, and $C$. These values implicitly depend on the density matrix of the quantum state, which can represent a coherent state, a squeezed state, or a superposition of coherent states. Furthermore, the expectation value of any operator $\boldsymbol{O}$,  denoted in the Schr\"odinger picture, is given by $\langle \boldsymbol{O} \rangle = \text{Tr}(\boldsymbol{\rho} \boldsymbol{O})$, where $\boldsymbol{\rho}$ denotes the density matrix. This leads to an important question: How does spin squeezing evolve as the density matrix changes over time, due to relaxation processes? Addressing this question requires analyzing the relationship between relaxation mechanisms and their influence on the squeezing parameter, which ultimately affects the quantum correlations, crucial for spin squeezing.

\section{Relaxation formalism} \label{Relaxation}

At the onset of establishing the theoretical framework for describing relaxation, several physical models have been proposed for spin systems with $I > 1/2$, introduced by J. Van Kranendonk \cite{van1954theory}, Yosida and Moriya \cite{yosida1956effects}, Kondo and Yamashita \cite{kondo1959nuclear}, and Kochelaev \cite{kochelaev1960longitudinal}. 
However, a suitable approach to describe relaxation for spin particles of more than two energy levels is through Redfield theory \cite{redfield1957theory}. Redfield theory is considered more complete because it utilizes concepts from quantum mechanics emplying the density matrix formalism. This approach takes into account the relaxation rates among the density matrix elements with the same coherence order, and fluctuations, due to any external field or particle surrounding the quantum system, are mathematically described and represented by spectral densities \cite{Ruben2008Relaxa, CONSUELOLEAL202372, leal2024dynamic}. Applying the Redfield approach, a differential equation is derived for each element of the density matrix, resulting in a system of $2I + 1 - n_{c}$ differential equations for each coherence order $n_{c}$  \cite{CONSUELOLEAL202372}. Therefore, the equation of motion for the density operator reads
\begin{eqnarray}
\frac{d\widetilde{\boldsymbol{\rho }}\left( t\right) }{dt} &=&i\left[ \widetilde{\boldsymbol{\rho }}\left( t\right) ,\widetilde{ \boldsymbol{\mathcal{H}}}_{NMR}^{SS}\left(t\right) \right] \nonumber \\ 
&-& \int\limits_{0}^{\infty }\left[ \left[ \widetilde{\boldsymbol{\rho }}\left(t\right) ,\widetilde{\boldsymbol{\mathcal{H}}}_{pert}\left( t^{\prime }\right) \right] ,
\widetilde{\boldsymbol{\mathcal{H}}}_{pert}\left( t\right) \right] dt^{\prime } \label{Redfield} \text{,}\quad
\end{eqnarray}
where $\widetilde{\boldsymbol{\mathcal{O}}}$  denotes any operator $\boldsymbol{\mathcal{O}}$ in the Interaction picture, $\widetilde{\boldsymbol{\mathcal{H}}}_{pert}$ represents the energy interaction between the quantum system and the fields or particles characterizing the environment. Furthermore, the double commutator operator helps with the characterization of the fluctuation effects due to the energy contributions of the perturbation Hamiltonian, meaning that if the Hamiltonian at time $t^{\prime}$ is strictly the same at  time $t$, fluctuation is not occurring and the equation of motion returns to its ideal isolated regime.  The  Hamiltonian that describes the fluctuations of the system studied in the interaction picture along the time is: $\widetilde{\boldsymbol{\mathcal{H}}}_{pert}\left( t\right)=\widetilde{\boldsymbol{\mathcal{H}}}_{Q}(t)$. Therefore, since the perturbed time-dependent Hamiltonian is the Hamiltonian that describes the quadrupole moment interacting with electric field gradients around the nucleus position of the space at $\boldsymbol{r}=(x,y,z)$, the Hamiltonian can be denoted as \cite{CONSUELOLEAL202372}
\begin{equation}
 {\boldsymbol{\mathcal{H}}}_{Q}(t) =\sum_{n=-2}^{n=2} F_{n}\left(\textbf{r}\left( t\right)\right)\boldsymbol{Q}^{(-n)} \text{.} \label{CompleteQuadrupolarHamiltonianToModelRelaxation}
\end{equation}

 The functions $F_{n}\left(\textbf{r}(t)\right)$ are denoted in terms of the second derivatives of any electric potential $V(\textbf{r}(t))$ at the nucleus position denoted by $\boldsymbol{r}(t)$ of the space and at any time $t$, and are generated by any charge distribution around the nucleus. They are determined as
\begin{eqnarray}
\!\!\!\!F_{\pm 2}\left(\textbf{r}\left( t\right)\right)  &=&\frac{V_{xx}\left(\textbf{r}\left( t\right)\right) \pm 2iV_{xy}\left(\textbf{r}\left( t\right)\right) -V_{yy}\left(\textbf{r}\left( t\right)\right) }{2\sqrt{6}}\text{, } \label{SecondRankF2} \\
\!\!\!\!F_{\pm 1}\left(\textbf{r}\left( t\right)\right)  &=&\frac{2V_{xy}\left(\textbf{r}\left( t\right)\right) \pm 2iV_{xy}\left(\textbf{r}\left( t\right)\right) }{\sqrt{6}}%
\text{,} \label{SecondRankF1} \\
\!\!\!\!F_{0}\left(\textbf{r}\left( t\right)\right)  &=&\frac{V_{zz}\left(\textbf{r}\left( t\right)\right) }{2}\text{.} \label{SecondRankF0}
\end{eqnarray}
Also, $\boldsymbol{Q}^{(-n)}$ represent $n$-rank tensor operators and these are expressed in terms of spin angular momentum operator as follows
\begin{eqnarray}
\mathbf{Q}^{\left( \pm 2\right) } &=&\frac{\sqrt{6}}{2}\frac{eQ}{2I\left(
2I-1\right) }\mathbf{I}_{\pm }^{2}\text{, }\label{Qope1} \\ 
\mathbf{Q}^{\left( \pm 1\right) } &=& \mp\frac{\sqrt{6}}{2}\frac{eQ}{2I\left(
2I-1\right) }\left( \mathbf{I}_{z}\mathbf{I}_{\pm }+\mathbf{I}_{\pm }\mathbf{I}_{z}\right) \text{,} \label{Qope2}\\ 
\mathbf{Q}^{\left( 0\right) } &=&\frac{eQ}{2I\left( 2I-1\right) }\left( 3\mathbf{I}_{z}^{2}-\mathbf{I}^{2}\right) \text{,} \label{Qope3}
\end{eqnarray}
where $e$ and $Q$ represent the elemental charge and the nuclear quadrupole moment, respectively. Futhermore, each term of the equation of motion in Eq.(\ref{Redfield}) has a physical interpretation in such a way that, for instance the first term represents the dynamics of an isolated system. In contrast, the second term describes the dynamics of the system under the effects of the interaction with a Markovian
environment, phenomenologically introduced, due to the fluctuations in the electric field gradients that generate the quadrupolar coupling. 

We now consider the energy contribution arising from the quadrupolar coupling, Eq.(\ref{CompleteQuadrupolarHamiltonianToModelRelaxation}), described in terms of the electric field gradients or second derivatives of an electric  potential function, Eqs.(\ref{SecondRankF2}-\ref{SecondRankF0}), and the $n$-rank tensor operators, Eqs.(\ref{Qope1}-\ref{Qope3}). After some mathematical procedures, it is possible to demonstrate that there is an equation of motion for each density matrix element \cite{Abragam1994, blicharski1999nmr, CONSUELOLEAL202372}
  \begin{equation}
\frac{d\left\langle \mathbf{T}_{l,m}\right\rangle }{dt}
\!=\!-C_{Q}\!\sum_{p=-2}^{2}(-1)^{p} J_{p}\! \left\langle 
\left[ \mathbf{Q}^{\left( p\right) },\left[ \mathbf{Q}^{\left( -p\right) },\mathbf{T}_{l,m}\right] \right]\right\rangle, \label{EqCq} 
\end{equation}
where $J_{p}=J\left( p\omega _{I}\right) $ denote the spectral density functions. In addition, the parity property is represented by $J\left( p\omega _{I}\right) =J\left(- p\omega _{I}\right) $, fulfilling the even function condition \cite{Kova2018,CONSUELOLEAL202372}. According to Eq.(\ref{EqCq}), the elements of the density matrix are determined in terms of the irreducible tensor operators $\mathbf{T}_{l,m}$, where $l$ and $m$ represent the rank and the order (coherence order) of the tensor element of an irreducible tensor operator basis, respectively. Here, the constant parameter $C_{Q}$ reads 
\begin{equation}
C_{Q}=\frac{9}{10} \frac{1}{(2I(2I-1))^{2}} \left(\frac{eQV_{zz}(r)}{\hbar } \right)^{2} \left( 1+\frac{\eta^2}{3}\right)\text{,}
\end{equation}
depends on the quadrupolar coupling and $\eta$ denotes the asymmetry parameter with $0 \leq \eta \leq 1$. 
 
 \section{Relaxation simulation} \label{simulation}
 
 The primary characteristic of Eq.(\ref{EqCq}) lies in constructing systems of linear differential equations. Afterward, we  evaluate the double commutator using matrix algebra properties and apply methods to solve differential equations. For the spin systems $I=3/2$ and $I=7/2$, the equations which describe the dynamics of each element of the density matrix were presented in Refs. \cite{Ruben2008Relaxa} and \cite{CONSUELOLEAL202372}, respectively. In our manuscript, we use the parameters already discussed to simulate the dynamics of the squeezing parameter under relaxation effects - Table.\ref{Table1}.

\begin{table}[h!]
\begin{tabular}{|c|c|c|}
\hline
 Parameters  & Sodium ($^{23}$Na) & Cesium ($^{133}$Cs) \\ \hline
$J_{0}$  &  $14$ ns    &   $590$ ns    \\ \hline
$J_{1}$  &   $4$ ns  &    $27$ ns  \\ \hline
$J_{2}$  &  $3.4$  ns  & $1.28$ ns   \\ \hline
$C_{Q}$  &  $1.2 \times 10^{10}$ Hz$^{2}$   & $9.9 \times 10^{6}$ Hz$^{2}$      \\ \hline
$\omega_{Q}/2\pi$  &  $16700$ Hz  & $5970$ Hz       \\ \hline
\end{tabular}
\caption{Parameter values used to simulate the dynamics of the squeezing parameter under relaxation effects for spins $3/2$ and $7/2$, considering sodium and cesium nuclei, respectively. } \label{Table1}
\end{table}

An important step in studying relaxation dynamics is to establish the steady state. In this work, we describe the steady state using the following density matrix
\begin{equation}
\boldsymbol{\rho}_{eq}=\frac{\boldsymbol{I}_{z}+I\boldsymbol{1}}{\text{Tr}({\boldsymbol{I}_{z}+I\boldsymbol{1}}) }  \text{,}    \label{rhoeq}
\end{equation}
where $I$ is the spin quantum number, $3/2$ or $7/2$. Furthermore, this choice ensures that the density operator trace is equal to $1$, which is consistent with the definition of the density matrix for a spin coherent state, since it represents a pure state.

We performed numerical simulations considering a lyotropic liquid crystal system immersed in a strong static magnetic field, $\boldsymbol{B}_{0}$.  Sodium dodecyl sulfate (SDS) and cesium pentadecafluorooctanoate (CS-PFO) were chosen to represent quadrupolar nuclei with spins $I=3/2$ and $I=7/2$, respectively.
We considered the quadrupolar Hamiltonian from Eq.(\ref{HNMRSS}) evolving under relaxation dynamics with the fluctuations also driven by the quadrupolar Hamiltonian. In other words, relaxation occurs because of fluctuations in the electric field gradient around the quadrupolar nucleus. The time intervals $\tau_{k}$ chosen to simulate relaxation dynamics were multiples of the inverse of the quadrupolar frequency, $2\pi/\omega_{Q}=\nu_{Q}^{-1}$, with $ \tau_{k} \in \left[k-1,k \right]\nu_{Q}^{-1}$ and $k=1,11, 21, \cdots,1001$.

\begin{figure}[t]
\begin{center}
\includegraphics[width=0.48\textwidth]{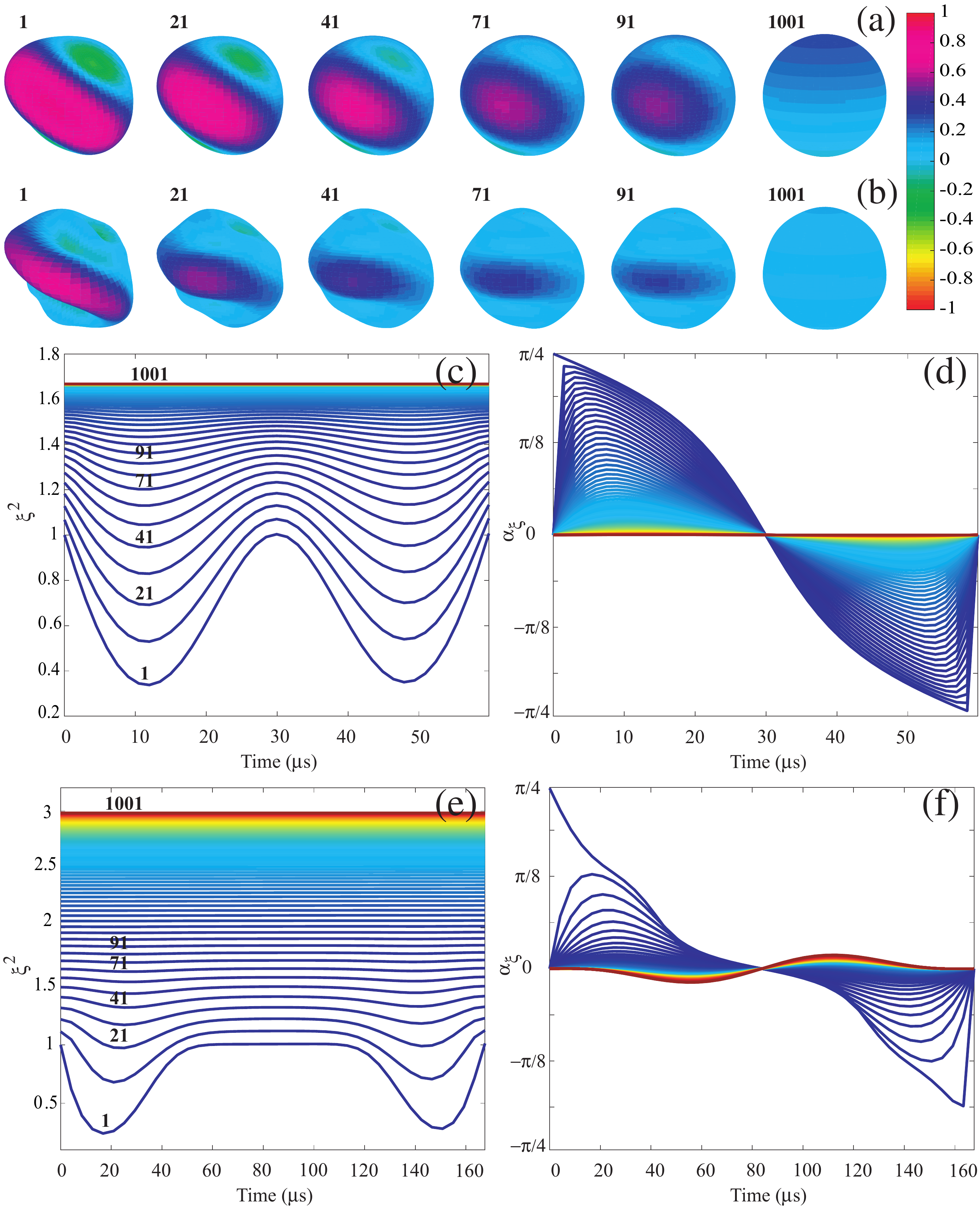}
\end{center}
\vspace{-0.6cm}
\caption{(Color online) (a)-(b) Theoretical Wigner quasi-probability distribution functions for spin systems $I=3/2$ and $I=7/2$, respectively, computed from a density matrix at time values $\protect\tau_{k}$ where $k=1,21,41,71,91,$ and $1001$. The spin coherent state $\left| \protect\zeta\left(\protect\pi/2,0\right)\right\rangle$ evolves under the Hamiltonian $ \boldsymbol{\mathcal{H}}^{SS}_{NMR}$. The dynamics of spins are analyzed during the time intervals $\tau_{k}$ with $ \tau_{k} \in \left[k-1,k \right]\protect\nu_{Q}^{-1}$. The distributions are generated according to the first minimum of the squeezing parameter as indicated in (c)-(e). Figures (c)-(d) and (e)-(f) correspond to the dynamics of the squeezing parameter and squeezing angle under relaxation for $I = 3/2$ and $I = 7/2$, respectively.}\label{fig:WignerTodos}
\end{figure}

The Wigner quasi-probability distribution in a generalized phase space of angular momentum was applied as a visual geometrical tool to elucidate the squeezing generation, and its disappear during the relaxation evolution. The quantum dynamics of the spin system was decoded from the density matrix evolved at time values $\tau_{k}$, where $k = 1, 21, 41, 71, 91,$ and $1001$, is presented. In both spin systems, the spin coherent state is 
 squeezed in the $yz$ plane, as shown in Figs.\ref{fig:WignerTodos}(a)-(b). Figs.\ref{fig:WignerTodos}(c)-(d) show the behavior of the squeezing parameter and angle during the relaxation dynamics for the $I=3/2$ spin system. For the $I=7/2$ case, these dynamics are shown in Figs.\ref{fig:WignerTodos}(e)-(f). In both cases, two minima are observed in the squeezing parameter throughout the full evolution. To generate the Wigner quasi-probability distributions, the first minimum was selected because it corresponds to the instant of maximum state squeezing. These minima are particularly significant because they indicate the instants of the time when the squeezing is most pronounced.

\section{Heisenberg uncertainty in the spin context} \label{uncertainty}

The Heisenberg uncertainty relation \cite{Heisenberg1927} is the most appropriate formal expression to characterize whether there is any correlation between a pair of quantum operators. This remarkable result led to the introduction of extensions in the quantum approach such that in 1929, when H. P.  Robertson demonstrated an equivalent mathematical relation considering two arbitrary observables $\boldsymbol{O}$ and $\boldsymbol{P}$  \cite{robertson1929uncertainty, trifonov2002generalizations} given by
\begin{equation}
 (\Delta \boldsymbol{O})^{2}\cdot (\Delta \boldsymbol{P})^{2} \geq \frac{1}{4} \vert \langle \left[ \boldsymbol{O},\boldsymbol{P}\right]   \rangle \vert^{2}         \text{,} \label{Rober}
 \end{equation}
where $\left[ \boldsymbol{O},\boldsymbol{P}\right] = \boldsymbol{O}\boldsymbol{P} - \boldsymbol{P}\boldsymbol{O}$ denotes the commutator of $\boldsymbol{O}$ and $\boldsymbol{P}$. This inequality establishes that the product of the variances of two observables is bounded by the expectation value of their commutator.

As shown in Figs.\ref{fig:WignerTodos}(a)-(b), the spin coherent state achieves the squeezing shape of the distribution probability on the $yz$ plane. Within this framework, the Heisenberg uncertainty relation can be analyzed by considering the spin operators $\boldsymbol{O}\equiv \boldsymbol{I}_{y}$ and $\boldsymbol{P}\equiv \boldsymbol{I}_{z}$, where $\left[ \boldsymbol{I}_{i},\boldsymbol{I}_{j}\right]=i\epsilon_{ijk}\boldsymbol{I}_{k}$ with $\epsilon_{ijk}$ denoting the Lévi-Cività symbol \cite{gross2012spin}
\begin{equation}
 \Delta \boldsymbol{I}_{y} \cdot\Delta \boldsymbol{I}_{z} \geq \frac{1}{2} \vert \langle \boldsymbol{I}_{x}  \rangle \vert         \text{.} \label{IyIz}
 \end{equation}

As we can see, the angle $\alpha_{\xi}$ shown in Figs.\ref{fig:WignerTodos}(d)-(f) is not always constant during the relaxation process; 
 the squeezing effect appears with the slope variation during the time evolution. It is necessary to calculate an appropriate linear combination of projections of the spin angular momentum operators $\boldsymbol{I}_{z}$ and $\boldsymbol{I}_{y}$.
To introduce a pair of angular momentum operators, we consider the unitary vector $\boldsymbol{u}$ in a three-dimensional real space $\mathbb{R}^{3}$  (see Tables $1$ and $2$ on page $7$ in Ref. \cite{zahn1979Book}) given by
\begin{equation}
\boldsymbol{u}=\left( u_{x},u_{y},u_{z}\right) =\left( \cos
\varphi \sin \vartheta ,\sin \varphi \sin \vartheta ,\cos \vartheta \right) \text{.}
\end{equation}%
The projection of the spin operator vector considering the
unitary vector on the $\mathbb{R}^{3}$ space, is denoted by%
\begin{equation}
\boldsymbol{I}_{p}=u_{x}\boldsymbol{I}_{x}+u_{y}\boldsymbol{I}_{y}+u_{z}\boldsymbol{I}_{z}\text{.}\label{Ip}
\end{equation}

In the squeezing formalism, the angle of squeezing $\alpha_{\xi}$ must be appropriately 
 characterized to match the geometric arrangement on the $yz$ plane  and computed from the physical meaning corresponding to the average values and variance of angular momentum operators. If $\varphi =\frac{\pi }{2}$  and $\vartheta =\alpha_{\xi} +\frac{\pi}{2}$, the unitary vector can be defined as
\begin{equation}
\boldsymbol{u}=\left( 0,\cos \alpha_{\xi} ,-\sin
\alpha_{\xi} \right) =\left( u_{x},u_{y},u_{z}\right) \text{,}
\end{equation}%
and substituting onto the expression of the angular momentum operator of Eq.(\ref{Ip})
\begin{equation}
\boldsymbol{I}_{p}=\boldsymbol{I}_{y}\cos \alpha_{\xi}  -\boldsymbol{I}_{z}\sin \alpha_{\xi}  \text{.}\label{Ipp}
\end{equation}%

To introduce another spin angular momentum operator, compatible with the previous $\boldsymbol{I}_{p}$, we define an orthogonal unitary vector also in
$yz$ plane
\begin{equation}
\boldsymbol{u}^{\perp }=\left( \cos \varphi \cos \vartheta ,\sin \varphi \cos \vartheta
,-\sin \vartheta \right) \text{.}
\end{equation}%
We assume the same values for the geometrical angle as previously identified, $\varphi =\frac{\pi }{2}$ and $\vartheta =\alpha_{\xi} +\frac{\pi }{2}$, such that the unitary orthogonal vector reads
\begin{equation}
\boldsymbol{u}^{\perp }=\left( 0,-\sin \alpha_{\xi}
,-\cos \alpha_{\xi} \right) \text{.}
\end{equation}%
By considering similar vector properties, the other compatible spin angular momentum operator is obtained as 
\begin{equation}
\boldsymbol{I}_{m }=-\boldsymbol{I}_{y}\sin \alpha_{\xi} -\boldsymbol{I}_{z}\cos \alpha_{\xi} \text{.}\label{Im}
\end{equation}%

Finally, the variance of both operators $\boldsymbol{I}_{p}$ and $\boldsymbol{I}_{m}$ can be computed using the linear combination of the angular momentum operators $\boldsymbol{I}_{y}$ and $\boldsymbol{I}_{z}$ employing Eq.(\ref{Ip}) and (\ref{Im})
\begin{equation}
 \Delta \boldsymbol{I}_{p} \cdot\Delta \boldsymbol{I}_{m} \geq \frac{1}{2} \vert \langle \boldsymbol{I}_{x}  \rangle \vert         \text{.} \label{IpIm}
 \end{equation}

\begin{figure}[h]
\begin{center}
\includegraphics[width=0.48\textwidth]{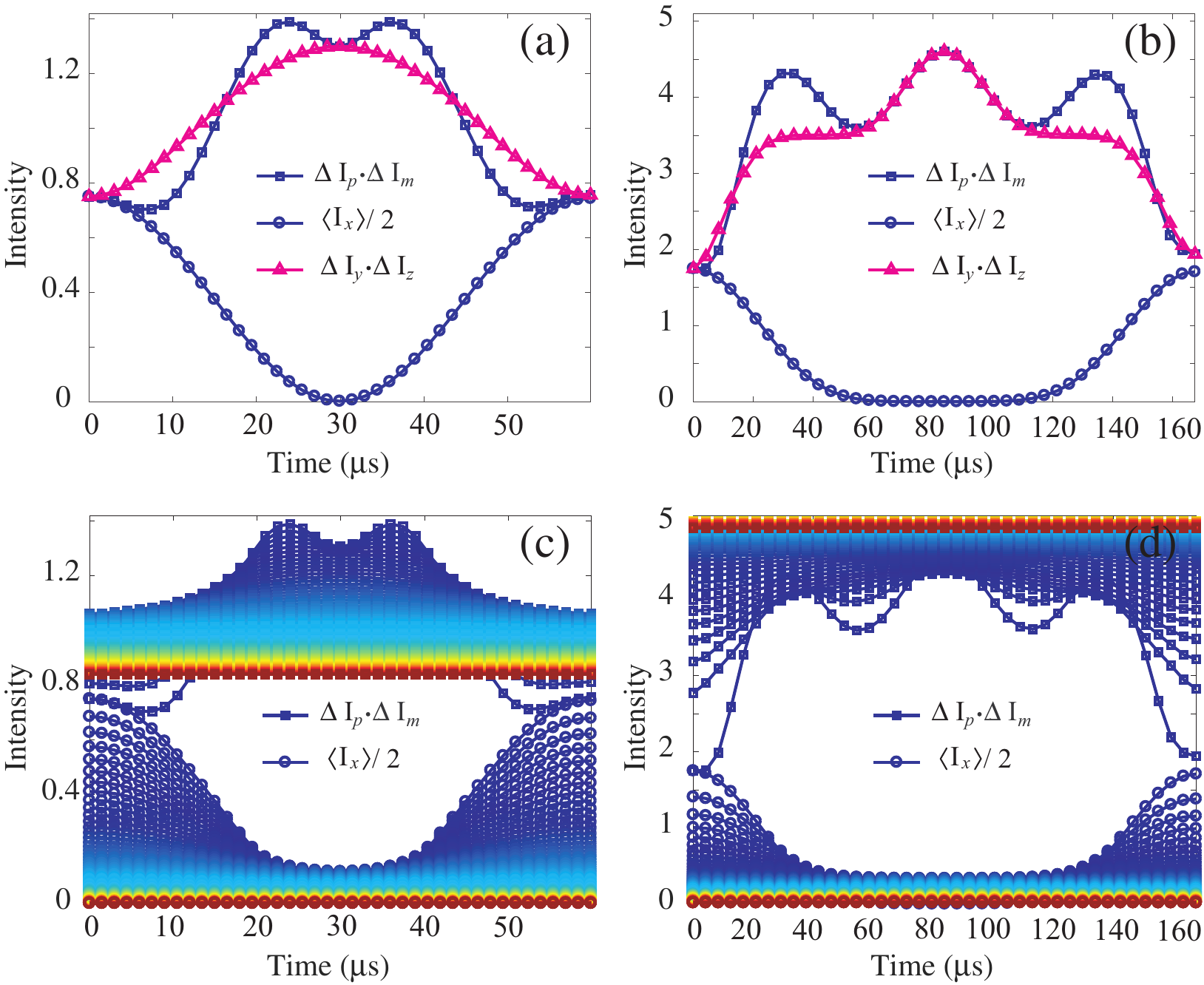}
\end{center}
\vspace{-0.6cm}
\caption{(Color online) The dynamics of the Heisenberg uncertainty relations, $\Delta \boldsymbol{I}_{p}\cdot \Delta \boldsymbol{I}_{m}$ and $ \Delta \boldsymbol{I}_{y}\cdot \Delta \boldsymbol{I}_{z}$, are compared to the average value of $\frac{\langle \boldsymbol{I}_{x}  \rangle}{2}$, for spin systems with $I=3/2$ and $I=7/2$. Figures (a) and (c) correspond to the spin-$3/2$, while figures (b) and (d) correspond to the spin-$7/2$ system. In (a) and (b), the time interval is $ \tau_{k} \in \left[0,k\protect\nu_{Q}^{-1} \right]$, where $k=1$. In (c) and (d), the dynamics are analyzed over multiple time intervals, $\tau_{k} \in \left[k-1,k \right]\protect\nu_{Q}^{-1}$ for $k=11, 21, \cdots, 1001$. }\label{fig:Medio}
\end{figure}

The Heisenberg uncertainty has different behavior for the variance of the pair of non-Cartesian angular momentum operators, $\Delta \boldsymbol{I}_{p}\cdot \Delta \boldsymbol{I}_{m}$. The same is true for the pair of Cartesian angular momentum operators, $\Delta \boldsymbol{I}_{y}\cdot \Delta\boldsymbol{I}_{z}$  in the time interval for the squeezing process. In Figs.\ref{fig:Medio}(a)-(b), both quantities are pictorially represented by blue-square and magenta-triangle symbols, respectively. The product of each pair of variance introduces a unique physical interpretation. The physical quantity associated with the pair of Cartesian angular momentum operators demonstrates greater sensitivity to the quantum coherences for the Schr\"odinger's cat state as shown by the magenta-triangle symbols in Figs.\ref{fig:Medio}(a)-(b).

\begin{figure}[t]
\begin{center}
\includegraphics[width=0.48\textwidth]{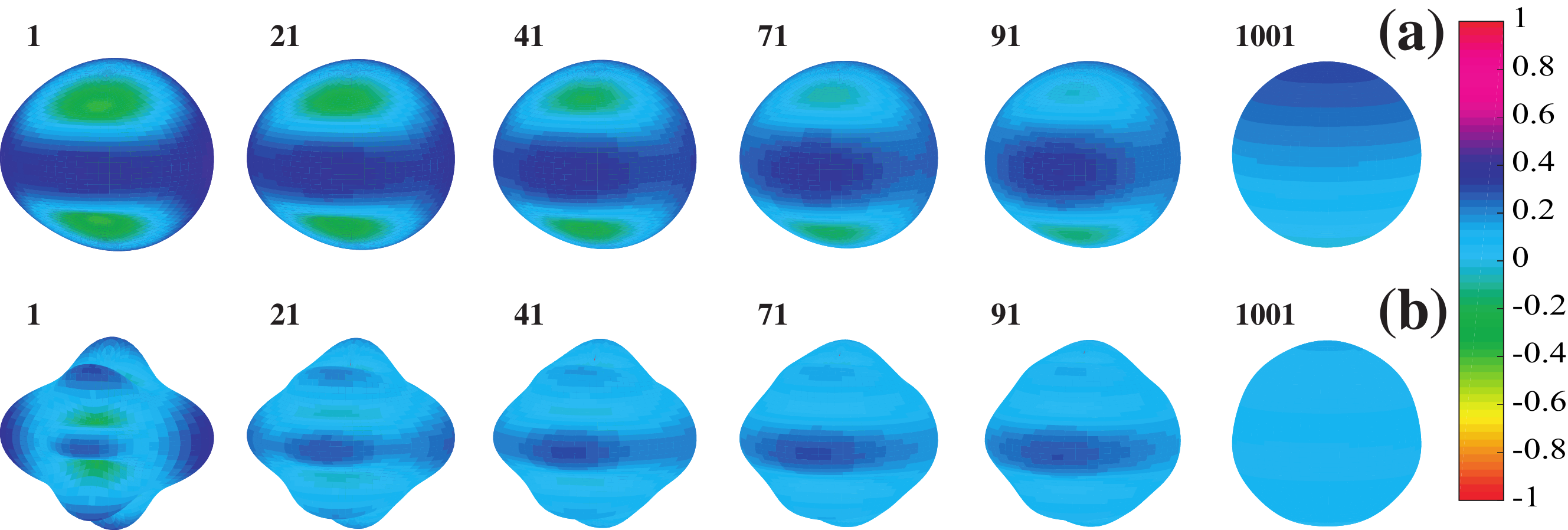}
\end{center}
\vspace{-0.6cm}
\caption{(Color online) (a)-(b) Theoretical Wigner quasi-probability distribution functions for spin systems with $I=3/2$ and $I=7/2$, respectively, are shown. These distributions are computed from the  density matrix at specific time values $\protect\tau_{k}$, where $k=1,21,41,71,91,$ and $1001$, corresponding to the evolution of the Schr\"odinger's cat state. The spin coherent state $\left| \protect\zeta\left(\protect \pi/2,0\right)\right\rangle$ evolves under the Hamiltonian $\boldsymbol{\mathcal{H}}^{SS}_{NMR}$, with the time interval $\tau_{k}=\frac{1}{2}\nu_{Q}^{-1}$ representing the generation of the Schr\"odinger's cat state.  }\label{fig:Gato}
\end{figure}

 Figs.\ref{fig:Gato}(a)-(b) show the theoretical Wigner quasi-probability distribution functions for spin $I=3/2$ and $I=7/2$, respectively.
The sensitivity of the distribution function highlights its enhanced compatibility with the Heisenberg uncertainty principle in effectively capturing quantum correlation effects during the generation of the Schr\"odinger's cat state. As the system evolves, the inequality gradually transits close to an equality regime. In contrast, the pair of non-Cartesian angular momentum operators, represented by blue-square symbols, exhibits sensitivity in capturing the quantum correlations throughout the dynamics of the spin system under scrutiny. This includes both the squeezing effect and the generation of the Schr\"odinger's cat state. In particular, for the pairs of non-Cartesian and Cartesian angular momentum operators, the inequalities converge at early time, which corresponds to the Schr\"odinger's cat state generation.

\begin{figure}[h]
\begin{center}
\includegraphics[width=0.48\textwidth]{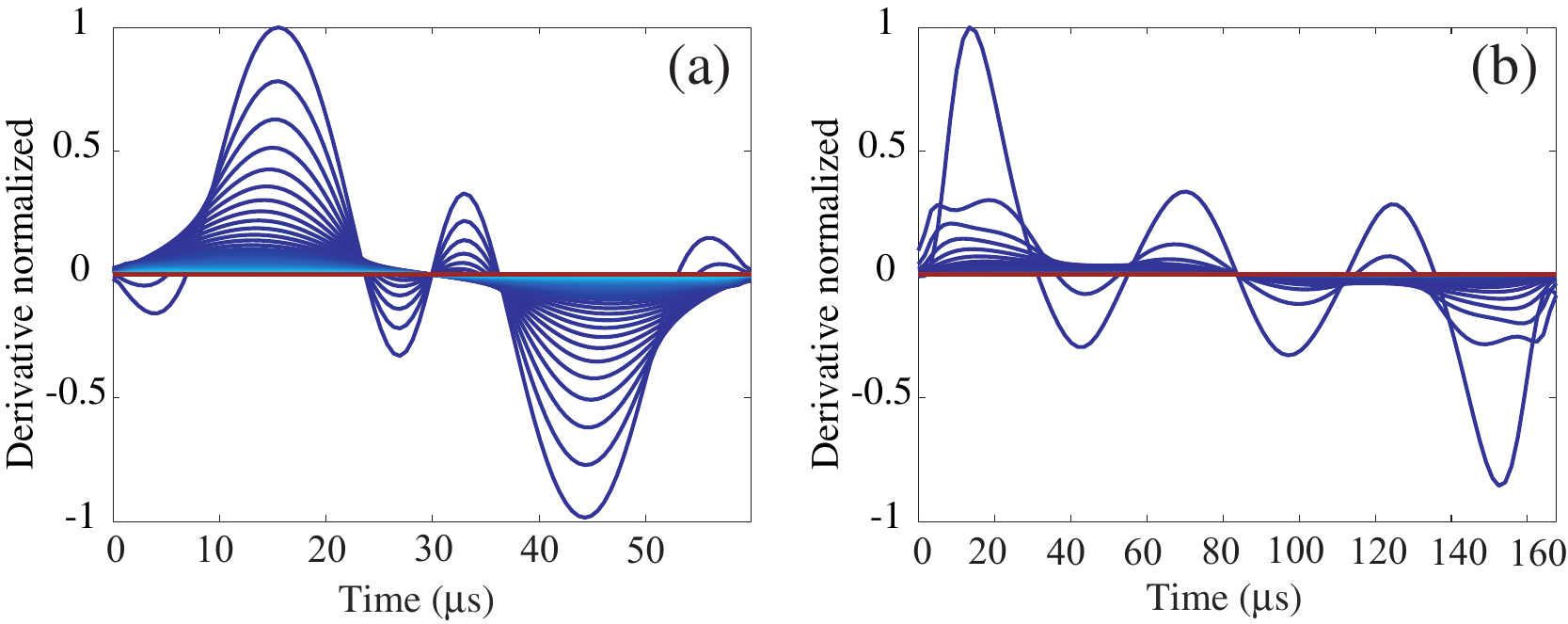}
\end{center}
\vspace{-0.6cm}
\caption{(Color online) Normalized derivatives of the product of variances for non-Cartesian angular momentum operators,  revealing the rate of change due to dynamics of quantum decoherence under the relaxation effects. (a) Spin system with $I = 3/2$; (b) Spin system with $I = 7/2$.  }\label{fig:Derivada}
\end{figure}

Considering decoherence under relaxation, the derivative of the product of variances for the non-Cartesian angular momentum operators (Figs.\ref{fig:Medio}(c) and \ref{fig:Medio}(d)) is calculated to investigate such behavior during the dynamics. This approach enables the evaluation of the rate of change between each pair of points. The results of this analysis are presented in Figs.\ref{fig:Derivada}(a) and (b), corresponding to the cases of spin $I=3/2$ and $I=7/2$, respectively. During the initial stages of evolution ($k = 1, 11, \ldots, 51$), the spin-$7/2$ system exhibits stronger quantum coherence, as shown in Figs.\ref{fig:Medio}(d) and (b), although this coherence fast dissipates. In contrast, the spin-$3/2$ system, while displaying weaker coherence amplitude, maintains the coherence for a longer period, persisting until approximately $k = 91$. This behavior is closely related to the spectral density parameter $J_{0}$ (see Table \ref{Table1}). This parameter influences the relaxation dynamics, particularly for zero- and high-order coherences. In the cesium nuclei, $J_{0}$ is significantly higher one order of magnitude greater than in the sodium nuclei, which causes faster decoherence in the spin-$7/2$ system.

\section{Bounds at the thermal equilibrium of the system} \label{upper}

At thermal equilibrium, the density matrix $\boldsymbol{\rho}_{eq}$ is well defined and described by Eq.(\ref{rhoeq}). In this context, we can predict an upper bound for the squeezing parameter $\xi_{eq}^{2}$. Under the same conditions, the angle $\alpha_{eq}$, which characterizes the geometric orientation of the squeezing, is zero. Consequently, the product of variances $(\Delta \boldsymbol{I}_{p} \cdot\Delta \boldsymbol{I}_{m})_{eq}$ and $(\Delta \boldsymbol{I}_{y} \cdot\Delta \boldsymbol{I}_{z})_{eq}$ becomes equivalent. An equation for both quantities can be established using the inductive method, a mathematical approach that begins with specific examples and extends  to general conclusions. This method is based on the principle that if a statement holds for a specific case and remains valid for a sufficiently large number of similar cases, it can be considered true for all cases. In that sense, this method allows us to find mathematical expressions for the squeezing parameter and the product of variances, expressed by the equations below

\begin{equation}
\xi_{eq}^{2}= \frac{2(I+1)}{3}\text{,} \quad
(\Delta \boldsymbol{I}_{y} \cdot\Delta \boldsymbol{I}_{z})_{eq}= \frac{(I+1)\sqrt{(2I-1)I}}{3\sqrt{3}}\text{.} \label{Retas}
\end{equation}

\begin{figure}[h]
\begin{center}
\includegraphics[width=0.48\textwidth]{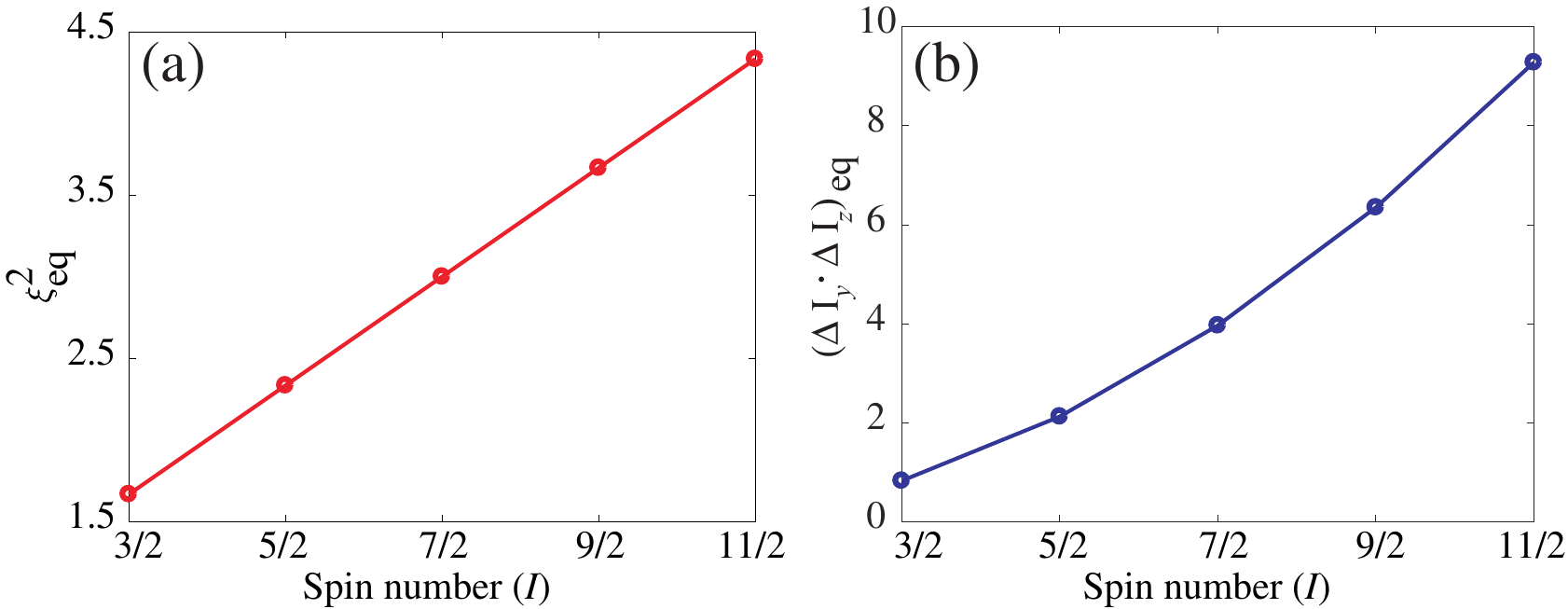} 
\end{center}
\vspace{-0.6cm}
\caption{(Color online) (a) Red dots represent the squeezing parameter at thermal equilibrium for each spin quantum number.
(b) Blue dots correspond to the product of variances under the same conditions.
The red and blue lines indicate the best fit to the data.}\label{fig:Reta}
\end{figure}

Fig.~(\ref{fig:Reta}) illustrates how the squeezing parameter and the product of the variances of two spin angular momentum operators depend on the spin quantum number in the thermal equilibrium regime. In Fig.\ref{fig:Reta}(a), the red dots represent the values of the squeezing parameter for different spin quantum numbers, while Fig.\ref{fig:Reta}(b) shows the corresponding product of variances (blue dots). The red and blue lines denote the best linear fits to each dataset. These results offer valuable insights with respect to the evolution of spin squeezing and uncertainty in quadrupolar spin systems, which allows analytical predictions.

For the specific cases presented in Figs.\ref{fig:WignerTodos}(c)-(e), the upper bound at thermal equilibrium is $1.67$ and $3.0$ for the squeezing parameter. Similarly, in Figs.\ref{fig:Medio}(c)-(d), the upper bound for the product of variances reaches $0.84$ and $4.0$, respectively. In all cases, these values are in agreement with the constraint given by Eq.(\ref{Retas}).

 Finally, an additional property of the non-Cartesian operators ${\boldsymbol{I}}_{p}$ and ${\boldsymbol{I}}_{m}$ preserves the macroscopicity measure for a quantum system \cite{frowis2012, gupta2024}, as introduced within the information-theoretic framework \cite{nimmrichter2013}.  The macroscopicity property, as demonstrated by the relative quantum Fisher information $N_{eff}$, is proportional to the variance of the measurement operator $\boldsymbol{O}$ evaluated at the quantum state described by the density matrix  $\boldsymbol{\rho}\left(t\right)$. The macroscopicity parameter can be expressed as
\begin{equation}
N_{eff}^{O}\left(\boldsymbol{\rho}\left(t\right)\right)=\frac{2}{{I}}\left( \left\langle \boldsymbol{O}^{2}\right\rangle -\left\langle \boldsymbol{O}\right\rangle ^{2}\right) \text{.}
\end{equation}

\begin{figure}[t]
\begin{center}
\includegraphics[width=0.483\textwidth]{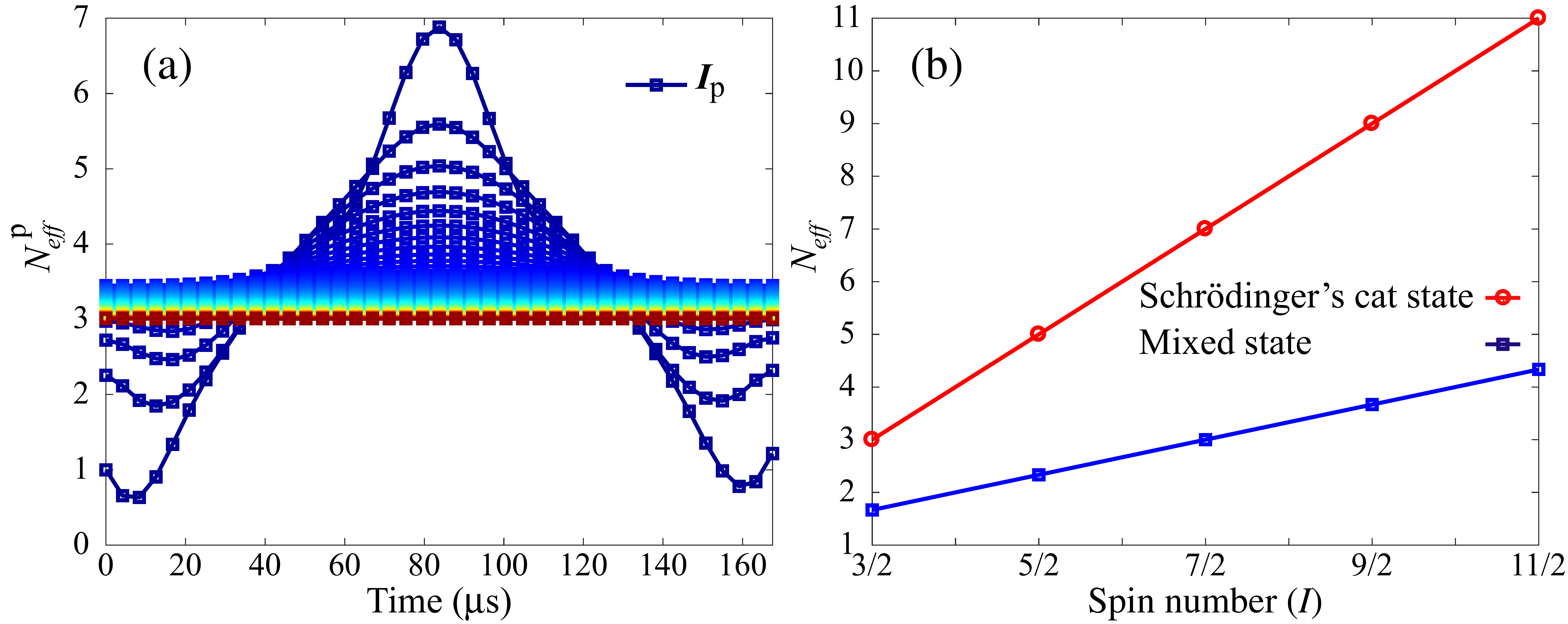} 
\end{center}
\vspace{-0.6cm}
\caption{(Color online) (a) Dynamics of the macroscopicity parameter $N_{eff}^{p}$ considering the angular momentum operator  ${\boldsymbol{I}}_{p}$ and spin $I = 7/2$. (b) Macroscopicity measure  $N_{eff}^{p}$ with a time interval $\tau=\frac{1}{2}\nu_{Q}^{-1}$, for the Schr\"odinger's cat state (red square symbols) and for the thermal equilibrium mixed state (blue square symbols).}\label{fig:Macroscopicity}
\end{figure}

Based on the definitions of the operators $\boldsymbol{I}_p$  and $\boldsymbol{I}_m$, given by Eqs.(\ref{Ipp}) and (\ref{Im}), respectively, the degree of superposition of the measurement operator reaches its maximum at the time when the system attains the  Schr\"{o}dinger's cat state, as also observed for its canonical counterpart $\boldsymbol{I}_y$. Instead, by considering the spin system with $I=7/2$ and the master equation that describes the relaxation dynamics, one can monitor the macroscopicity parameter $N_{eff}^{p}$ associated with the angular momentum operator ${\boldsymbol{I}}_{p}$ (see Fig.\ref{fig:Macroscopicity}(a)). Moreover,  applying this macroscopicity quantifier, the system exhibits a monotonic decrease in size, from the maximum value $N_{eff}^{max}\left(\boldsymbol{\rho}_{Cat}\right)= 2I$ down to the minimum value $N_{eff}\left(\boldsymbol{\rho}_{eq}\right)=2\left( I+1\right) /3$,  at which the system reaches the thermal equilibrium state $\boldsymbol{\rho}_{eq}$, as defined in Eq.(\ref{rhoeq}). The amount of macroscopicity loss due to environmental  relaxation effects corresponds to $N_{eff}^{max}\left(\boldsymbol{\rho}_{Cat}\right)-N_{eff}\left(\boldsymbol{\rho}_{eq}\right) = 2\left( 2I-1\right) /3$. Both quantities - the maximum and minimum macroscopicity sizes - are  shown as functions of the spin number in Fig.\ref{fig:Macroscopicity}(b). 

In this analysis, macroscopicity acquires an interesting physical meaning, as it reveals that, as the spin value of a quantum system increases, the robustness of quantum properties remains resilient. This behavior arises from the fact that the rate of increase in macroscopicity surpasses the value attained at thermal equilibrium. The aforementioned aspects, along with several other parameters, are of particular interest for exploration in various physical contexts~\cite{nimmrichter2013,frowis2018}, as well as in promising recent experimental developments \cite{moreno-pineda2021,maylander2023,little2023}.

\section{Conclusions} \label{Conclusions}

In this work, we investigate the decoherence of spin squeezed states in NMR quadrupolar systems with spin quantum numbers $I = 3/2$ and $I = 7/2$, incorporating  relaxation effects. The spin squeezing arises from the interaction between the nuclear quadrupole moment and the local electric field gradients. By employing the spin angular momentum representation within the one-axis twisting mechanism, we establish a formal description of the nonlinear operators governing spin squeezing, providing a theoretical basis for quantum control of this phenomenon while accounting for relaxation dynamics.

Furthermore, we introduce a set of non-Cartesian angular momentum operators whose variance products catch the quantum correlations underlying the spin dynamics. A key result of this study is the derivation of an upper bound for the squeezing parameter and the Heisenberg uncertainty product at thermal equilibrium, generalized for arbitrary spin quantum numbers. These results deepen the understanding of quantum coherence and correlations in quadrupolar spin systems, offering a foundation for further theoretical and experimental advancements in quantum information processing and precision metrology based on nuclear spin ensembles.

 %%%%%%%%%%%%%%%%%%%%%%%%%%%%%%%%%%%%%%%%%%%%%%%%%%%%%%%%%%%%%%%%%%%%%%%%%%%%%%%%%%%%%%%%%%%%%%%%%%
 
\section{Acknowledgments}
A. C. L Auccaise knowledges the CNPq and Fundação Araucária for financial support under project  $174469/2024-1$.

%%%%%%%%%%%%%%%%%%%%%%%%%%%%%%%%%%%%%%%%%%%%%%%%%%%%%%%%%%%%

\end{document}